
\input phyzzx
\hsize=6truein
\def\TITLEPAGE{\frontpagetrue}
\def\CALT#1{\hbox to\hsize{\tenpoint \baselineskip=12pt
	\hfil\vtop{\hbox{\strut CALT-68-#1}
	\hbox{\strut DOE RESEARCH AND}
	\hbox{\strut DEVELOPMENT REPORT}}}}

\def\CALTECH{\smallskip
	\address{California Institute of Technology, Pasadena, CA 91125}}
\def\TITLE#1{\vskip 1in \centerline{\fourteenpoint #1}}
\def\AUTHOR#1{\vskip .5in \centerline{#1}}

\def\ABSTRACT#1{\vskip .5in \vfil \centerline{\twelvepoint \bf Abstract}
	#1 \vfil}
\def\ENDTITLEPAGE{\vfil\eject\pageno=1}

\def\sqr#1#2{{\vcenter{\hrule height.#2pt
      \hbox{\vrule width.#2pt height#1pt \kern#1pt
        \vrule width.#2pt}
      \hrule height.#2pt}}}

\def\section#1#2{
\noindent\hbox{\hbox{\bf #1}\hskip 10pt\vtop{\hsize=5in
\baselineskip=12pt \noindent \bf #2 \hfil}\hfil}
\medskip}

\def\underwig#1{	
	\setbox0=\hbox{\rm \strut}
	\hbox to 0pt{$#1$\hss} \lower \ht0 \hbox{\rm \char'176}}

\def\bunderwig#1{	
	\setbox0=\hbox{\rm \strut}
	\hbox to 1.5pt{$#1$\hss} \lower 12.8pt
	 \hbox{\seventeenrm \char'176}\hbox to 2pt{\hfil}}

\def\MEMO#1#2#3#4#5{
\frontpagetrue
\centerline{\tencp INTEROFFICE MEMORANDUM}
\smallskip
\centerline{\bf CALIFORNIA INSTITUTE OF TECHNOLOGY}
\bigskip
\vtop{\tenpoint
\hbox to\hsize{\strut \hbox to .75in{\caps to:\hfil}\hbox to 3.8in{#1\hfil}
\quad\the\date\hfil}
\hbox to\hsize{\strut \hbox to.75in{\caps from:\hfil}\hbox to 3.5in{#2\hfil}
\hbox{{\caps ext-}#3\qquad{\caps m.c.\quad}#4}\hfil}
\hbox{\hbox to.75in{\caps subject:\hfil}\vtop{\parindent=0pt
\hsize=3.5in #5\hfil}}
\hbox{\strut\hfil}}}

\def\frac#1#2{{#1\over#2}}

\tolerance=1000
\hfuzz=5pt
\TITLEPAGE
\CALT{1774}
\TITLE {Evaporation of Two Dimensional Black Holes}
\AUTHOR{S. W. Hawking\foot {Work supported in part by the U.S. Dept. of
Energy under Contract no. DEAC-03-81ER40050.}}
\CALTECH
\centerline {and}
\centerline {\it Department of Applied Mathematics and Theoretical Physics}
\centerline {\it University of Cambridge}
\centerline {\it Silver Street Cambridge CB3 9EW, UK}
\ABSTRACT{Callan, Giddings, Harvey and Strominger have proposed an
interesting two
dimensional model theory that allows one to consider black hole evaporation
in the semi-classical approximation.  They originally hoped the black hole
would evaporate completely without a singularity. However it has been shown
that the semi-classical equations will give a singularity where the dilaton
field reaches a certain critical value. Initially, it seems this singularity
will be hidden inside a black hole. However, as the evaporation proceeds, the
dilaton field on the horizon will approach the critical value but the
temperature and rate of emission will remain finite. These results indicate
either that there is a naked singularity, or (more likely) that the
semi-classical approximation breaks down when the dilaton field approaches the
critical value.}
\smallskip
\leftline {February 1992}
\ENDTITLEPAGE

\noindent {\bf Introduction}

Callan, Giddings, Harvey and Strominger (CGHS) [1] have suggested an
interesting two dimensional theory with a metric coupled to a dilaton field
and $N$ minimal scalar fields. The Lagrangian is
$$L = {1\over{2\pi}}\sqrt{-g}[e^{-2\phi}(R+4(\nabla\phi)^2+4\lambda^2) -
{1\over 2} \sum_{i=1}^N(\nabla f_{i})^2],$$
If one writes the metric in the form
$$
ds^2= e \sp {2\rho} dx_+ dx_-
$$
the classical field equations are
$$
\partial_{+}\partial_{-}f_{i}=0,
$$
$$2\partial_{+}\partial_{-}\phi-2\partial_{+}\phi\partial_{-}\phi -
{\lambda^2 \over 2} e^{2\rho}=\partial_{+}\partial_{-}\rho,$$
$$\partial_{+}\partial_{-}\phi-2\partial_{+}\phi\partial_{-}\phi -
{\lambda^2 \over 2} e^{2\rho}=0.$$

These equations have a solution
$$\phi =-b\log (-x_+x_-)-c-\log \lambda$$
$$\rho =-{1\over 2}\log (- x_+x_-)+\log {2b\over \lambda}$$
where $b$ and $c$ are constants and $b$ can be taken to be positive without
loss of generality. A  change of coordinates
$$
u\pm =\pm {2b\over \lambda}\log (\pm x_\pm )\pm {1\over \lambda}(c+ \log
\lambda)
$$
 gives a flat metric and a linear dilaton field
$$
\rho =0
$$
$$
\phi =-{\lambda\over 2}(u_+-u_-)
$$
 This solution is known as the linear dilaton. The solution
is independent of the constants $b $ and $c$ which correspond to freedom in
the choice of coordinates. Normally $b$ is taken to have the value
${\textstyle{1 \over 2}} $.

These equations also admit a solution
$$
\phi=\rho  -c = -{1\over 2}\log (M \lambda \sp {-1}-\lambda e^{2c}x_+x_-)
$$.
This represents a two dimensional black hole with horizons at
$x_{\pm}=0$ and singularities at $x_+x_-=M \lambda \sp {-2}e^{-2c}$. Note that
there is still freedom to shift the $\rho $ field on the horizon by a constant
and compensate by rescaling the coordinates $x_{\pm}$,  but there's nothing
corresponding to the freedom to choose the constant $b$. In terms of the
coordinates $u_{\pm}$ defined as before with $b={1\over 2}$
$$
\rho = -{1\over 2}\log (1-M \lambda \sp {-1} e^{-\lambda (u_+-u_-)})
$$
$$
\phi =-{\lambda \over 2}(u_+-u_-) -{1\over 2}\log (1-M \lambda \sp {-1}
e^{-\lambda (u_+-u_-)})
$$

This black hole solution is periodic in the imaginary time with period
$2\pi \lambda \sp {-1}$. One would therefore expect it to have a temperature
$$
T={ \lambda \over 2\pi}
$$
and to emit thermal radiation [2]. This is confirmed by CGHS. They considered a
black hole formed by sending in a thin shock wave of one of the $f \sb i $
fields from the weak coupling region (large negative $\phi $) region of the
linear dilaton. One can calculate the energy momentum tensors of the $f \sb i
$ fields, using the conservation and trace anomaly equations. If one imposes
the boundary condition that there is no incoming energy momentum apart from
the shock wave, one finds that at late retarded times $u_-$ there a steady
flow of energy in each $f \sb i $ field at the mass independent rate
$$
\lambda \sp 2 \over 48
$$

If this radiation continued indefinitely, the black hole would radiate an
infinite amount of energy, which seems absurd. One might therefore expect that
the back reaction would modify the emission and cause it to stop when the
black hole had radiated away its initial mass. A fully quantum treatment of
the back reaction seem very difficult even in this two dimensional theory. But
CGHS suggested that in the limit of a large number $N $ of scalar fields $f
\sb i $, one could neglect the quantum fluctuations of the dilaton and the
metric and treat the back reaction of the radiation in the $f \sb i$ fields
semi-classically by adding to the action a trace anomaly term
$$
{N\over 12}\partial_{+}\partial_{-}\rho.
$$

The evolution equations that result from this action are
$$
\partial_{+}\partial_{-}\phi=
(1-{N\over24}e^{2\phi})\partial_{+}\partial_{-}\rho,
$$
$$
2(1-{N\over 12}e^{2\phi})\partial_{+}\partial_{-}\phi=(1-{N\over
24}e^{2\phi})(4\partial_{+}\phi\partial_{-}\phi+\lambda^{2}e^{2\rho}).
$$
In addition there are two equations that can be regarded as constraints on the
data on characteristic surfaces of constant $x_\pm $
$$
(\partial_{+}^{2}\phi-2\partial_{+}\rho\partial_{+}\phi) ={N\over 24}e^{2\phi}
(\partial_{+}^{2}\rho- \partial_{+}\rho\partial_{+}\rho-t_{+}(x^{+})),
$$
$$
(\partial_{-}^{2}\phi-2\partial_{-}\rho\partial_{-}\phi) ={N\over 24}e^{2\phi}
(\partial_{-}^{2}\rho-\partial_{-}\rho\partial_{-}\rho-t_{-}(x^{-})),
$$
 where $t_{\pm}(x_{\pm})$ are determined by the boundary conditions in a manner
 that will be explained later.

Even these semi-classical equations seem too difficult to solve in closed form.
 CGHS suggested that a black hole formed from an $f $ wave would evaporate
completely without there being any singularity. The solution would approach
the linear dilaton at late retarded times $u_-$ and there would be no
horizons. They therefore claimed that there would be no loss of quantum
coherence in the formation and evaporation of a two dimensional black hole:
the radiation would be in a pure quantum state, rather than in a mixed state.

In [3] and [4] it was shown that this scenario could not be correct. The
solution would develop a singularity on the incoming $f $ wave at the point
where the dilaton field reached the critical value
$$
\phi \sb 0 =-{\textstyle{1 \over 2}} \log {N \over 12}
$$
This singularity will be spacelike near the $f$ wave [4]. Thus at least part of
the final quantum state will end up on the singularity, which implies that the
radiation at infinity in the weak coupling region will not be in a pure
quantum state.

The outstanding question is: How does the spacetime evolve to the future of the
 $f$ wave? There seem to be two main possibilities:

\item {1} The singularity remains hidden behind an event horizon. One can
continue an infinite distance into the future on a line of constant $\phi <
\phi \sb 0$ without ever seeing the singularity. If this were the case,  the
rate of radiation would have to go to zero.

\item {2} The singularity is naked. That is, it is visible from a line of
constant $\phi $ at a finite time to the future of the $f$ wave. Any evolution
of the solution after this would not be uniquely determined by the semi-
classical equations and the initial data. Indeed, it is likely that the point
at which the singularity became visible was itself singular and that the
solution could not be evolved to the future for more than a finite time.

In what follows I shall present evidence that suggests the semi-classical
equations lead to possibility 2. This probably indicates that the semi-
classical approximation breaks down as the dilaton field on the horizon
approaches the critical value.

\noindent {\bf Static Black Holes}

If the solution were  to evolve without a naked singularity, it would
presumably approach a static state in which a singularity was hidden behind an
event horizon. This motivates a study a study of static black hole solutions
of the semi-classical equations. One could look for solutions in which $\phi $
and $\rho $ depended only on a `radial' variable $\sigma =x_+-x_-$ but this
has the disadvantage that the black hole horizon is at $\sigma =-\inf$.
Instead it seems better to define the radial coordinate to be
$$
r^2=-x_+x_-
$$
The horizon is then at $r =0$ and the field equations for a static solution
are:
$$
\phi''+{1\over r} \phi'=\left( 1-{N \over 24} e^{2\phi} \right)
\left( \rho '' +{1\over r} \rho ' \right)
$$
$$
\left( 1-{N \over 12 }e^{2\phi} \right) \left( \phi '' +{1\over r} \phi '
\right) =2 \left( 1-{N \over 24} e^{2\phi} \right) \left( (\phi ')^2- \lambda
\sp 2 e^{2\rho} \right)
$$
 The boundary conditions for a regular horizon are
$$
\phi ' = \rho ' =0
$$
A static black hole solution is therefore determined by the values of $ \phi $
and $\rho $ on the horizon. The value of $\rho $ however can be changed by a
constant by rescaling the coordinates $x_\pm$. The physical distinct static
solutions with a horizon are therefore characterized simply by $\phi \sb h$,
the value of the dilaton on the horizon.

If $\phi \sb h>\phi \sb 0$, $\phi $ would increase away from the horizon and
would always be greater than its horizon value. This shows that to get a
static black hole solution that is asymptotic to the weak coupling region of
the linear dilaton, $\phi \sb h $ must be less than the critical value $\phi
\sb 0$. One can then show that both $\phi $ and $ \rho $ must decrease with
increasing $r$. This means the back reaction terms proportional to $N$ will
become unimportant. For large $r$ one can therefore approximate by putting
$N=0$. This gives
$$
\phi = \rho - (2 b-1)\log r -c
$$
$$
\phi'' +{1\over r}\phi ' =2( (\rho'-(2b-1) r^{-1})^2-\lambda \sp 2 e^{2\rho})
$$
Asymptotically these have the solution
$$
\rho =-\log r+\log {2b\over \lambda}-{ K+L\log r \over r^{4b}}+...
$$
where $b,c,K,L$ are parameters that
determine the solution. The parameters $b$and $ c$ correspond to the
coordinate freedom in the linear dilaton that the solution approaches at large
$r$. The parameter $L$ does not appear in the black hole solutions. If it is
zero, the parameter $K$ can be related to the ADM mass $M$ of the solution.
The effects of the back reaction terms proportional to $N$ will affect only
the higher order terms in $r^{-1}$.

For $\phi \sb h<< \phi \sb 0$, the back reaction terms will be small at all
values of $r$ and the solutions of the semi-classical equations will be almost
the same as the classical black holes. So
$$
\phi \sb 0=- {\textstyle{1 \over 2}} \log {M\over \lambda}
$$

Consider a situation in which a black hole of large mass $(M>>N \lambda /12)$
is created by sending in an $f$ wave. One could approximate the subsequent
evolution by a sequence of static black hole solutions with a steadily
increasing value of $\phi $ on the horizon. However, when the value of $\phi $
on the horizon approaches the critical value $\phi \sb 0$, the back reaction
will become important and will change the black hole solutions solutions
significantly. Let
$$
\phi =\phi \sb 0+\bar \phi, \rho =\log \lambda +\bar\rho
$$
Then $N$ and $\lambda $ disappear and the equations for static black
holes become
$$
\bar \phi '' +{1\over r} \bar \phi ' ={1 \over 2} \left(
2-e^{2\bar \phi} \right) \left( \bar \rho '' +{1\over r} \bar \rho ' \right)
$$
$$
\left( 1-e^{2\bar \phi} \right) \left( \bar \phi '' +{1\over r} \bar
\phi ' \right) =\left( 2-e^{2\bar \phi} \right) \left( (\bar \phi
')^2-e^{2\bar \rho} \right)
$$

As the dilaton field on the horizon approaches the critical value $\phi\sb 0$,
the term $(1-e^{2\bar \phi})$ will approach $2\epsilon $, where $\epsilon
=\phi \sb 0-\phi \sb h $. This will cause the second derivative of $\bar \phi
$ to be very large until $\bar \phi ' $ approaches $-e^{\bar \rho \sb h}$ in a
coordinate distance $\Delta r$ of order $4\epsilon $. By the above equations,
$\rho ' $ approaches $-2e^{\bar \rho \sb h}$ in the same distance. A power
series solution and numerical calculations carried out by Jonathan Brenchley
confirm that in the limit as $\epsilon $ tends to zero, the solution tends to
a limiting form $\bar \phi \sb c, \bar \rho \sb c $.

The limiting black hole is regular everywhere outside the horizon, but has a
fairly mild singularity on the horizon with $R $ diverging like $r^{-1}$. At
large values of $r$, the solution will tend to the linear dilaton in the
manner of the asymptotic expansion given before. One or both of the constants
$K$ and $L$ must be non zero, because the solution is not exactly the linear
dilaton. Fitting to the asymptotic expansion gives a value
$$
b \sb c\approx 0.4
$$

If the singularity inside the black hole were to remain hidden at all times,
as in possibility (1) above, one might expect that the temperature and rate of
evolution of the black hole would approach zero as the dilaton field on the
horizon approached the critical value. However, this is not what happens. The
fact that the black holes tend to the limiting solution $\bar \phi \sb c,
\bar \rho \sb c $ means that the period in imaginary time will tend to $ 4
\pi b \sb c \over \lambda $. Thus the temperature will be
$$
T \sb c={\lambda\over 4 \pi b\sb c}
$$
The energy momentum tensor of one of the $f_i$ fields
can be calculated from the conservation equations. In the $x_\pm$ coordinates,
they are:
$$
\left\langle T_{++}^f \right\rangle =-{1 \over 12}(\partial _+\bar
\rho \partial _+\bar \rho -\partial _+^2\bar \rho +t_+(x_+)),
$$
$$
\left\langle
T_{--}^f \right\rangle =-{1 \over 12}(\partial _- \bar \rho \partial _- \bar
\rho -\partial _- ^2\bar \rho +t_-(x_-))
$$
where $t_\pm (x_\pm )$ are chosen to satisfy the boundary conditions on the
energy momentum tensor. In the case of a black hole formed by sending in an
$f$ wave, the boundary condition is that the incoming flux $ \left\langle
T_{++}^f \right\rangle $ should be zero at large $r$. This would imply that
$$
t_+={1\over 4x_+^2}
$$
The energy momentum tensor would not be regular on the
past horizon, but this does not matter as the physical spacetime would not
have a past horizon but would be different before the $f$ wave.

On the other hand, the energy momentum tensor should be regular on the future
horizon. This would imply that $ t_-(x_-)$ should be regular at $x_-=0$.
Converting to the coordinates $u_{\pm}$, one then would obtain a steady rate
$$
\lambda \sp 2 \over 192 b \sb c \sp 2
$$
 of energy outflow in each $f$ field at late retarded times $u_-$.
\vfil \eject

\noindent {\bf Conclusions}

The fact that the temperature and rate of emission of the limiting black hole
do not go to zero, establishes a contradiction with the idea that the black
hole settles down to a stable state. Of course, this does not tell us what the
semi-classical equations will predict, but it makes it very plausible that
they will lead either to a naked singularity, or to a singularity that spreads
out to infinity at some finite retarded time.

The semi-classical evolution of these two dimensional black holes, is very
similar to that of charged black holes in four dimensions with a dilaton field
[5]. If one supposes that there are no fields in the theory that can carry
away the charge, the steady loss of mass would suggest that the black hole
would approach an extreme state. However, unlike the case of the
Reissner-Nordstom solutions, the extreme black holes with a dilaton have a
finite temperature and rate of emission. So one obtains a similar
contradiction. If the solution where to evolve to a state of lower mass but
the same charge, the singularity would become naked.

There seems no way of avoiding naked singularity in the context of the
semi-classical theory. If spacetime is described by a semi-classical Lorentz
metric, a black hole can not disappear completely without there being some
sort of naked singularity. But there seem to be zero temperature non radiating
black holes only in a few cases. For example, charged black holes with no
dilaton field and no fields to carry away the charge.

What seems to happening is that the semi-classical approximation is breaking
down in the strong coupling regime. In convential general relativity, this
breakdown occurs only when the black hole gets down to the Planck mass. But in
the two and four dimensional dilatonic theories, it can occur for macroscopic
black holes when the dilaton field on the on the horizon approaches the
critical value. When the coupling becomes strong, the semi-classical
approximation will break down. Quantum fluctuations of the metric and the
dilaton could no longer be neglected. One could imagine that this might lead
to a tremendous explosion in which the remaining mass energy of the black hole
was released. Such explosions might be detected as gamma ray bursts.

Even though the semi-classical equations seem to lead to a naked singularity,
one would hope that this would not happen in a full quantum treatment. Quite
what it means not to have naked singularities in a quantum theory of gravity
is not immediately obvious. One possible interpretation is the no boundary
condition [6]: spacetime is non singular and without boundary in the Euclidean
regime. If this proposal is correct, some sort of Euclidean wormhole would
have to occur, which would carry away the particles that went in to form the
black hole, and bring in the particles to be emitted. These wormholes could be
in a coherent state described by alpha parameters [7]. These parameters might
be determined by the minimumization of the effective gravitational constant
$G$ [7,8,9]. In this case, there would be no loss of quantum coherence if a
black hole were to evaporate and disappear completely. Or the alpha parameters
might be different moments of a quantum field $\alpha $ on superspace[10]. In
this case there would be effective loss of quantum coherence, but it might be
possible to measure all the alpha parameters involved in the evaporation of a
black hole of a given mass. In that case, there would be no further loss of
quantum coherence when black holes of up to that mass evaporated.

\bigskip

I was greatly helped by talking to Giddings and Stominger
who were working along similar lines. I also had useful
discussions with Hayward, Horowitz and Preskill. This work
was carried out during a visit to Cal Tech as a Sherman
Fairchild Scholar.

\noindent {\bf References}

\item {1.} Callan, C.G., Giddings, S.B., Harvey, J.A.,
Strominger, A. Evanescent Black Holes UCSB-TH-91-54.
\item {2.} Hawking, S.W. Particle Creation by Black Holes, Commun.
Math. Phys. 43,199 (1975).
\item {3.} Banks, T., Dabholkar, A., Douglas, M.R., O'Loughlin, M. Are
Horned Particles the Climax Of Hawking Evaporation?
RU-91-54.
\item {4.} Russo, J.G., Susskind, L., Thorlacius, L. Black Hole
Evaporation in 1+1 Dimensions SU-ITP-92-4.
\item {5.} Garfinkle, D., Horowitz, G.T., Strominger, A. Charged
Black Holes in String Theory, Phys. Rev D 43, 3140.
\item {6.} Hartle, J.B., Hawking, S.W. Wave Function of the
Universe Phys. Rev. D28, 2960-2975 (1983).
\item {7.} Coleman, S. Why There Is Nothing Rather Than Something:
A Theory Of The Cosmological Constant Nucl. Phys. B310
(1988), 643.
\item {8.} Preskill, J.  Wormholes In Spacetime And The Constants
Of Nature.  Nucl. Phys. B323 (1989), 141.
\item {9.} Hawking, S.W. Do Wormholes Fix The Constants Of
Nature? Nucl. Phys. B335,155-165 (1990).
\item {10.} Hawking, S.W. The Effective Action For Wormholes.  Nucl.
Phys. B363, 117-131 (1991).
\vfill
\end